\documentclass[a4paper,11pt]{article}

\usepackage{jheppub_modified} 
\usepackage{graphicx}   
\usepackage[T1]{fontenc} 
\usepackage{slashed}
\usepackage{feynmp}
\usepackage{epsfig}
\usepackage{tikz}
\usepackage{enumerate}
\newcounter{exno}
\newcommand{\ex}{\stepcounter{exno}\theexno}
\title{Physics Beyond the Standard Model}
\subheader{2014 BUSSTEPP Lectures}

\author{Ben Gripaios}
\affiliation{Cavendish Laboratory,\\JJ Thomson Avenue,
\\ Cambridge, 
CB3 0HE,
United Kingdom.\\ \\ \today}

\emailAdd{gripaios@hep.phy.cam.ac.uk}

\begin{document} 
\maketitle
\flushbottom

\acknowledgments
I thank my collaborators and many lecturers who
have provided inspiration over the years; some of their excellent
notes can be found in the references. The references are intended to
provide an entry point into the literature, and so are necessarily
incomplete; I apologize to those left out.
For errors and comments, please contact
me by e-mail at the address on the front page.

\section{Avant propos}
On the one hand, giving 4 lectures on physics beyond the Standard
Model (BSM) is an impossible task, because there is so much that one
could cover. On the other hand, it is easy, because it can be summed
up in the single phrase: There must be something, but we don't know
what it is! This is, of course, tremendously exciting. It has to be
said that we also know a great deal about what BSM physics {\em isn't}, because of
myriad experimental and theoretical constraints. Whether this is to be
considered good news or bad news is somewhat subjective. At least it keeps us honest.

My aim in these lectures is to give a flavour of the field of BSM
physics today, with an attempt to focus on aspects which are not so
easy to find in textbooks. I have tried to make the lectures
introductory and to dumb them down as much as
possible. I apologize in advance to those who feel that their
intelligence is being insulted. 

A number of shorter derivations are left as exercises. These are
numbered and
indicated by `(exercise $n$)' where they appear.

\section{Notation and conventions}
As usual, $\hbar = c= 1$, and our metric is mostly-minus: $\eta^{\mu
  \nu} = \mathrm{diag} (1,-1,-1,-1)$.

We also need to decide on conventions for fermions. This is somewhat involved, and
I don't have the time to do it properly. Let me
at least lay down the rules of the game.
I assume that the reader is familiar with Dirac (4-component) fermions, for which
a set of gamma matrices, satisfying
$\{\gamma^\mu , \gamma^\nu\} \equiv \gamma^\mu 
\gamma^\nu + \gamma^\nu  \gamma^\mu = 2\eta^{\mu \nu} $, is
\begin{gather}
\gamma^\mu = \begin{pmatrix}  0 & \sigma^\mu \\
  \overline{\sigma}^{\mu} & 0\end{pmatrix},
\end{gather} 
where $\sigma^\mu = (1, \sigma^i)$, $  \overline{\sigma}^{\mu} = (1,
-\sigma^i)$, and $\sigma^i$ are the usual 2 x 2 Pauli matrices:
\begin{gather}
\sigma^1 = \begin{pmatrix}  0 & 1 \\
  1& 0 \end{pmatrix},
\sigma^2 = \begin{pmatrix}  0 & -i \\
  i& 0 \end{pmatrix},
\sigma^3 = \begin{pmatrix}  1 & 0 \\
  0& -1 \end{pmatrix}.
\end{gather} 
Hence,
\begin{gather}
\gamma^5 \equiv i \gamma^0 \gamma^1 \gamma^2 \gamma^3 = \begin{pmatrix}  -1 & 0 \\
 0 & 1\end{pmatrix}.
\end{gather} 

Dirac fermions are fine for QED and QCD, but they are not what appears
in the SM.\footnote{A gauge theory containing
  Dirac fermions is called {\em vector-like}; otherwise it is called {\em chiral}.}
A Dirac fermion carries a 4-dimensional, {\em reducible}
representation (henceforth `rep') of the Lorentz group $SO(3,1)$.\footnote{One way to see that it is reducible is to consider the complex form of
the Lie algebra which is isomorphic to the complex form of
$SU(2)\times SU(2)$. (A complex form is obtained by taking
the original Lie algebra, which is a real vector space, and promoting
the coefficients in linear superpositions from $\mathbb{R}$ to
$\mathbb{C}$.) The Dirac fermion corresponds to the
$(2,1) \oplus (1,2)$ rep of $SU(2)\times SU(2)$, which is
reducible. Note that while $(2,1)\oplus (1,2)$ is a unitary rep of
$SU(2)\times SU(2)$, the Dirac fermion does not carry a unitary rep of
$SO(3,1)$, because of factors of $i$ that appear in taking the
different real forms. This is why things like $\overline{\psi} \psi$
appear rather than $\psi^\dagger \psi$.}
We can, therefore, assign fermions to carry either of the 2, 2-d
irreducible reps (henceforth `irreps') that are carried by a 4-d Dirac fermion. These are called
Weyl fermions. The
2 2-d  irreps are inequivalent. We'll let
one of the irreps be carried by a 2-component object $\psi_{\alpha}$, with
$\alpha \in \{1,2\}$, and the other
be carried by a different
2-component object $\overline{\chi}^{\dot{\alpha}}$, with
$\dot{\alpha} \in \{1,2\}$. We put a bar on $\overline{\chi}$ and a dot on
$\dot{\alpha}$ to indicate that they are different sorts of object to $\psi$
and $\alpha$, respectively. We also put one index upstairs and one
index downstairs, for reasons that will become clear. 

These two representations are conjugate, meaning that the
complex conjugate of one is equivalent (via a similarity transformation) to the other. We thus make the
definitions
\begin{gather}
\overline{\psi}_{\dot{\alpha}} \equiv (\psi_\alpha)^*, \; \chi^\alpha
\equiv (\overline{\chi}^{\dot{\alpha}})^*.
\end{gather}

We now introduce conventional definitions for raising and lowering indices, {\em
  viz.}
\begin{align} \label{eq:weylconj}
\psi_\alpha &\equiv \epsilon_{\alpha \beta} \psi^\beta, \;
\psi^\beta \equiv \epsilon^{\beta \gamma} \psi_\gamma, \\
\overline{\psi}_{\dot{\alpha}} &\equiv \epsilon_{\dot{\alpha}
  \dot{\beta}} \overline{\psi}^{\dot{\beta}}, \;
\overline{\psi}^{\dot{\beta}} \equiv \epsilon^{\dot{\beta}
  \dot{\gamma}} \overline{\psi}_{\dot{\gamma}},
\end{align}
where $\epsilon_{\alpha \beta} \equiv \begin{pmatrix} 0 & 1 \\ -1 & 0\end{pmatrix}$.
Note that these imply, {\em e.g.}, $\epsilon_{\alpha
  \beta}\epsilon^{\beta \gamma} = \delta^{\gamma}_\alpha$, so
$\epsilon^{\alpha \beta} \equiv \begin{pmatrix} 0 & -1 \\ 1 &
  0\end{pmatrix}$. Similarly, the above relations imply
$\epsilon_{\alpha \beta} = \epsilon_{\dot{\alpha} \dot{\beta}}$.

The point of all this notation (for which you can thank Van der
Waerden), is that, just like for 4-vectors in relativity, anything with all undotted
and dotted 
indices (separately) contracted upstairs and downstairs in pairwise fashion is a Lorentz invariant. (Look in a QFT
book for a proof.)

This means that given just a single Weyl spinor, $\psi_\alpha$ we can
write a mass term in the lagrangian of the form $\psi^{\alpha}
\psi_\alpha$ (to which we should add the Hermitian conjugate to be
sure that the action is real). This is called a Majorana mass term,
and we can write it if $\psi$ transforms in a real
rep of any internal symmetry group (because the
product of a rep and its conjugate contains a singlet). For the same
reason, if we have two Weyl
spinors in conjugate reps $\psi_\alpha$  and $\chi_\alpha$, we can write a mass term of
the form $\psi^\alpha \chi_\alpha + \mathrm{h.\ c.}$. This is called a Dirac mass term.

To save drowning in a sea of indices, it is useful to define
$\psi^\alpha \chi_\alpha \equiv \psi \chi $ and
$\overline{\psi}_{\dot{\alpha}} \overline{\chi}^{\dot{\alpha}} \equiv
\overline{\psi} \overline{\chi}$. But note that, {\em e.g.}, $\psi_\alpha
\chi^\alpha = - \psi \chi$ (exercise \ex). Fortunately, since the fermions take
Grassman number values, we do have that $\psi \chi = \chi \psi$
(exercise \ex).

Going back to 4-component language, we can write a Dirac fermion as
$\Psi = \begin{pmatrix} \psi_\alpha &
  \overline{\chi}^{\dot{\alpha}}\end{pmatrix}^T$. A Majorana fermion can
be written as $\Psi = \begin{pmatrix} \psi_\alpha &
  \overline{\psi}^{\dot{\alpha}}\end{pmatrix}^T$, such that $\Psi =
\Psi^c$ (exercise \ex).

Lorentz-invariant terms can also be formed using the same rules with
the objects $(\sigma^\mu)_{\alpha \dot{\alpha}}$ and
$(\overline{\sigma}^\mu)^{ \dot{\alpha} \alpha}$, which appear in
$\gamma^\mu$.\footnote{Again, this is just group theory: a 4-vector
  corresponds to the $(2,2)$ irrep of $SO(3,1)$, and we can make something that
  transforms like a $(2,2)$ by taking the (tensor) product of $(1,2)$
  and $(2,1)$ irreps.} So we may write
terms $i \chi^\alpha \sigma^\mu_{\alpha \dot{\alpha}}
\partial_\mu \overline{\chi}^{\dot{\alpha}} \equiv i\chi
\sigma^\mu \partial_\mu \overline{\chi}$ and $i \overline{\psi}_{\dot{\alpha}}
\overline{\sigma}^{\mu \dot{\alpha} \alpha} \partial_\mu \psi_\alpha \equiv i
\overline{\psi} \overline{\sigma}^\mu \partial_\mu \psi$
and indeed the usual Dirac kinetic term produces the sum of these
(exercise \ex).

Since we can get from one rep to the other by taking the complex
conjugate, we can, w.\ l.\ o.\ g.\, assign all fermions to just one
irrep, which we take to be the undotted one.

\section{The Standard Model}
To go {\em beyond} the Standard Model (SM), we first must know something about the
SM itself. We define the SM as a gauge
quantum field theory with gauge symmetry $SU(3) \times SU(2) \times
U(1)$, together with matter fields comprising 15 Weyl fermions and one
complex scalar, carrying irreps of $SU(3) \times SU(2) \times
U(1)$. The fermions consist of 3 copies (the different families or
flavours or generations) of 5 fields, $\psi \in \{q,u^c, d^c, l, e^c\}$,
carrying reps of $SU(3) \times SU(2) \times
U(1)$ as listed in Table~\ref{tab:smreps}. The 
scalar field, $H$, carries the $(1,2,-\frac{1}{2})$ rep of $SU(3) \times SU(2) \times
U(1)$. The final part of the definition is that the lagrangian should
contain all terms up to dimension four, such that it is
renormalizable. 
\begingroup
\begin{table*}[ht]
\begin{center}
\begin{tabular}{c  c  c  c  }
\hline
\hline
Field &  $SU(3)_c $ & $SU(2)_L $   & $U(1)_Y $  \\
\hline
$q $ &3 &2 &$+\frac{1}{6}$ \\
$u^c$ & $\overline{3}$& 1& $-\frac{2}{3}$\\
$d^c$ & $\overline{3}$& 1& $+\frac{1}{3}$\\
$l $ &1 &2 & $-\frac{1}{2}$ \\
$e^c$ &1 & 1& $+1$\\
\hline
\end{tabular}
\end{center}
\caption{Fermion fields of the SM and their $SU(3)\times SU(2) \times
U(1)$ representations.\label{tab:smreps}}
\end{table*}
\endgroup

To see how elegant the SM is, we note that we can write the lagrangian
on a single line (just). It is, schematically,
\begin{gather}
\mathcal{L} = i \overline{\psi}_i \overline{\sigma}^\mu  D_\mu \psi_i
-\frac{1}{4} F^a_{\mu \nu} F^{a \mu \nu} + \lambda^{ij} \psi_i \psi_j
H^{(c)} + \mathrm{h.\ c.}+ |D_\mu H|^2 - V(H),
\end{gather}
where $i,j$ label the different families and $a$ labels the different
gauge fields. 
You might counter that the way I choose to write it is arbitrary (and
indeed the paradox of ``the greatest integer which cannot be described in fewer
  than twenty words'' is brought to mind). So let's explore it in more
  detail.
\subsection{The gauge sector}
The lagrangian is
\begin{gather}\label{eq:lgauge}
\mathcal{L}  = i \overline{\psi}_i \overline{\sigma}^\mu  D_\mu \psi_i-\frac{1}{4} F^a_{\mu \nu} F^{a \mu \nu},
\end{gather}
with 5 fermion irreps $\psi \in \{q,u^c,d^c,l,e^c\}$ and 3 copies of
each, corresponding to the 3 families. There are really 12 gauge
fields: 8 in an adjoint of $SU(3)$, 3 in an adjoint of $SU(2)$, and 1
for $U(1)$. The covariant derivative $D_\mu$
contains the gauge couplings $g_s, \, g,$ and $g^\prime$, with the gauge
group generators in the appropriate reps.
The fermion terms are invariant under a $U(3)^5$ global
symmetry. (Exercise \ex: show this. What is the global symmetry when the gauge
coupings are switched off?)
This part of the SM has been tested at the per mille level via charge
universality and tests of gauge boson
interactions ({\em e.g.} $\sigma (e^+ e^- \rightarrow W^+
W^-)$. Exercise \ex: draw the Feynman diagrams that contribute at
leading order to $\sigma (e^+ e^- \rightarrow W^+
W^-)$ in the SM).
\subsection{The flavour sector}
The lagrangian is
\begin{gather} \label{eq:lyuk}
\mathcal{L} = \lambda^u q H^c u^c + \lambda^d qH d^c + \lambda^e l H e^c +
\mathrm{h.\ c.}
\end{gather}
The $\lambda^i$ are 3 $3\times 3$ complex matrices (in family space)
and so it appears that there are a lot of free parameters, and also
the possibility of $CP$ violation (since a $CP$ transformation is
equivalent to interchanging the $\lambda^i$s with their complex
conjugates. Exercise \ex). But not all of these parameters are physical.
This follows from the fact that we are free to do
unitary rotations of the different fields without changing other terms
in the lagrangian. So, for example, there is a basis in which we can
write (exercise \ex)
\begin{gather} \label{eq:lyukd}
\lambda^u q H^c u^c + \lambda^d V qH d^c + \lambda^e l H e^c +
\mathrm{h.\ c.}
\end{gather}
where now all the $\lambda^i$ are diagonal, and $V$ is a $3\times 3$
unitary matrix, called the CKM matrix. Since this is the only
off-diagonal object in the lagrangian, it must contain all the
information about mixing of flavours in the SM. Very roughly, we find
that
\begin{gather}  \label{eq:ckm}
V \sim \begin{pmatrix} 1 & \lambda & \lambda^3 \\
\lambda & 1 & \lambda^2 \\
\lambda^3 & \lambda^2 & 1\end{pmatrix},
\end{gather}
where $\lambda \sim 0.2$.

To actually count the number of physical parameters, we need to be a
bit more careful.
In the quark sector, the Yukawa terms in (\ref{eq:lyuk}) break the $U(3)^3$ symmetry down to
the $U(1)_B$ corresponding to baryon number conservation. In the
lepton sector, $U(3)^2$ is broken to $U(1)_{L_e}\times
U(1)_{L_\mu} \times U(1)_{L_\tau}$, corresponding to conservation of
individual lepton family numbers. There is also the overall $U(1)_Y$
which is conserved.

Knowing the pattern of symmetry breaking, we can count the number of
physical parameters in, {\em e.\ g.}, the quark sector. There are two
complex $3 \times 3$ matrices in the quark sector, with a total of 18
real and 18 imaginary parameters (equivalently, there are 18 complex
phases).
But many of these parameters are unphysical, in the sense that they
can be removed using the unbroken `symmetries'.\footnote{They are not
  really symmetries, because they do not leave the lagrangian invariant!} An $N\times N$ unitary
matrix has $N^2$ parameters, of which $\frac{N(N-1)}{2}$ are real and
$\frac{N(N+1)}{2}$ are imaginary (for an Hermitian matrix, it is the
other way around. Exercise \ex: Prove these results.) So in the quark
sector, with $U(3)^3$ broken to $U(1)_B$, we have 9 unbroken real
parameters and 17 phases, meaning that there are $18-9=9$ physical
real parameters and $18-17=1$ physical phase. If we now refer to
(\ref{eq:lyukd}), we see that 6 of the real parameters are the quark
masses, so there must be 3 physical angles in the CKM matrix, and a
single phase. This phase is a source of $CP$-violation.

The flavour sector of the SM is tested at the per cent level, in many
experiments (observations of rare processes). As we shall see below,
this puts constraints on new physics that are much more stringent than
one might na\"{\i}vely guess.

\subsection{The Higgs sector}
The lagrangian is
\begin{gather}
\mathcal{L} = -\mu^2 H^\dagger H - \lambda (H^\dagger H)^2 .
\end{gather}
Until recently, this sector was hardly tested at all. But now, with
the discovery of the Higgs boson, it is being probed directly at
the ten per cent level.

So, ugly or not, the SM does an implausibly good job of describing the
data, reaching the per mille level in individual measurements and with
an overall fit (to hundreds of measurements) that cannot be denied:
the SM is undoubtedly correct, at least in the regime in which we are currently
probing it. 

\section{Miracles of the SM}
So far we have written down a definition of the SM, and argued that it
gives a compelling explanation of current data. This approach has
a major deficiency, which is that, in simply writing
it down, we completely overlook the heroic (and sometimes tragi-comic) struggles of our forefathers over many decades to arrive at it.

On the one hand, this is exactly the sort of youthful disrespect for
one's elders that leads to great advances in physics, and should be
encouraged as much as possible. But on the other hand, it means that
we miss certain features of the SM that are very special, and are
vital clues in our quest for the form of physics BSM. Let us discuss
some of them.

\subsection{Flavour}
Let's look in more detail at the flavour structure (for introductory lectures on this topic, see \cite{Grossman:2010gw,Gedalia:2010rj}). We have already
argued that there is a basis in which we can write the Yukawa couplings
as (in a matrix notation for flavour)
\begin{gather}
q \lambda^u  H^c u^c + q \lambda^d V H d^c + l \lambda^e  H e^c +
\mathrm{h.\ c.}
\end{gather}
To go from here to the mass basis after EWSB, all we need to do is a
rotation, $d \rightarrow V^\dagger d$, of
the $d$ quarks in $q = ( u \; d)^T$. This has the effect of making the gauge
interactions non-diagonal. In particular, we find charged current
interactions involving the $W^\pm$ of the form
\begin{gather}
g (\overline{u} \overline{\sigma} \cdot W^+ V^\dagger d + \overline{d}
V \overline{\sigma} \cdot W^- u).
\end{gather}
The neutral current interactions involving the $Z$ remain diagonal, however:
\begin{gather}
-g \overline{d} V \overline{\sigma} \cdot W^3 V^\dagger d = -g
\overline{d}  \overline{\sigma} \cdot W^3 d,
\end{gather}
since $VV^\dagger = 1$.
This is a key feature of the SM, which was motivated by the
experimental absence of FCNC. In fact, the absence of FCNC in the SM
(at tree-level -- we discuss loop interactions below)
also extends to the other neutral currents in the SM, {\em viz.} those
involving gluons, photons, or Higgs bosons. For the gluons and
photons, this arises simply because the corresponding gauge symmetries are
unbroken in the vacuum, and the couplings are universal. 
More prosaically, the coupling matrices are proportional to the
identity matrix, and so are diagonal in any basis. 
For the Higgs
boson, it arises because the couplings of the Higgs to fermion are
diagonal in the mass basis. This is because (in unitary gauge),
$H = \begin{pmatrix} 0 \\ v +h \end{pmatrix}$ and so we are
diagonalizing the same matrix for the fermion masses as for the
couplings of the fermions to the Higgs boson. But note that this would no longer necessarily be true in a
theory with more than 1 Higgs doublet, where there are extra Yukawa
coupling matrices in general, and the EW VEV is shared between the
different doublets (Exercise \ex: Show explicitly that $\exists$
tree-level FCNC in a 2 Higgs doublet model. How is this avoided in the
MSSM?).

The absence of tree-level FCNC mediated by $Z$ bosons is also a
special feature of the SM. It arises because all fields carrying the
same irrep of the unbroken $SU(3)_c\times U(1)_Q$ symmetry (which can mix with each
other) also carry the same irrep of the broken $SU(3)_c\times SU(2)_L \times
U(1)_Y$ symmetry. 
If this were not the case, the different irreps would, in general, have
different couplings to the $Z$. The matrix of couplings to the $Z$
would then be
diagonal in the interaction basis, but not proportional to the
identity matrix. The matrix would then
acquire off-diagonal
entries when we rotate to the mass basis.
For
example, the $d, s$ are both colour anti-triplets with charge
$-\frac{1}{3}$, and can mix, but they also all come from $SU(2)$
doublets with hypercharge $+\frac{1}{6}$, so there are no FCNC.
(Exercise \ex: before the charm quark was invented, it was thought that
the $s$ quark lived in an $SU(2)$ singlet. Show that this leads to
tree-level FCNC.) 

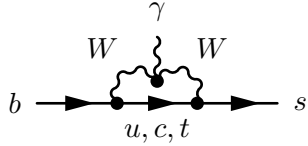
\begin{figure}
\centering 
\begin{fmffile}{diagram1} 
 \begin{fmfgraph*}(90,50)
  \fmfleftn{i}{1} \fmfrightn{o}{1} \fmftopn{t}{1}
   \fmflabel{$b$}{i1}
    \fmflabel{$s$}{o1}
\fmflabel{$\gamma$}{t1}
       \fmf{fermion}{i1,v1}
    \fmf{fermion,label=$u,,c,,t$}{v1,v2}
       \fmf{fermion}{v2,o1}
   \fmffreeze
   \fmf{boson,label=$W$,left=1}{v1,v3}
\fmf{boson,label=$W$,left=1}{v3,v2}
\fmf{boson}{v3,t1}
        \fmfdot{v1,v2,v3}
 \end{fmfgraph*}
 \end{fmffile} 
\caption{Diagram contributing to $b
\rightarrow s \gamma$ \label{fig:bsg}.}
\end{figure}

There are also remarkable suppressions of loop level processes.
Consider, for example the diagram contributing to the process $b
\rightarrow s \gamma$ in Fig.~\ref{fig:bsg}.
The internal quark can be any one of $u,c,t$ and so the amplitude is
proportional to
\begin{gather}
\sum_{i\in \{u,c,t\}} V_{ib} V^*_{is} f(\frac{m_i^2}{m_W^2})
\end{gather}
where $f$ is some function obtained by doing the loop
integral.\footnote{Note that these loop integrals are finite. This
  must be the case, because they generate operators in the low-energy effective
  lagrangian with four fermions, for which there are no counterterms
  available in the renormalizable SM.} Now,
suppose we do a Maclaurin expansion of $f$. The first term in the sum
then vanishes, by unitarity of the CKM matrix. At the next order, we
have terms that go like $\frac{m_u^2}{m_W^2}$ (which is tiny),
$\frac{m_c^2}{m_W^2}$ (which is small), and $\frac{m_t^2}{m_W^2}$
(which is certainly not small, but whose contribution is supressed by
$V_{ib} V^*_{is}$; the same is true at higher order in
$\frac{m_t^2}{m_W^2}$). (Exercise \ex: show that the latter 2 contributions
are roughly the same size for the analogous process $s \rightarrow d
\gamma $. Is this true for all
processes?) We thus find that these loop diagrams feature a
GIM \cite{Glashow:1970gm} suppression. This suppression is
in addition to the factor of $(\frac{1}{4\pi})^2$ that comes from the
fact that we have to do a loop integral: $\int \frac{d^4p}{(2\pi)^4}
f(p^2) = (\frac{1}{4\pi})^2\int xdx f(x)$ (exercise \ex: show this).
Overall, the SM contribution is of size
$\frac{1}{(4\pi)^2}\frac{m_c^2}{m_W^2}\frac{1}{m_W^2}$ compared to
a generic new physics contribution with mass scale
$\Lambda$ and $O(1)$ couplings  of $\frac{1}{\Lambda^2}$. Thus the
latter are greatly enhanced and the bounds on $\Lambda$ are typically way above
$m_W$
(given that the SM contributions give a good fit to the data).
In fact, they reach as high as $10^5$ GeV or so.
\subsection{$CP$-violation}
The peculiar structure of the flavour sector in the SM, and the fact
that $CP$-violation resides in the CKM matrix, imply that there are
also suppressions of $CP$-violating processes that do not occur in
generic models BSM. To see this, note that if there had only been two
generations of quarks in the SM, there would be no physical
$CP$-violating parameter in the flavour sector (exercise \ex). This means
that any process which violates $CP$ in the SM must involve all 3
quark generations. For similar reasons, $CP$-violation cannot occur if
any of the masses are degenerate in either the up or down sector, or
if any of the 3 mixing angles is 0 or $\frac{\pi}{2}$: all of these
situations increase the symmetry in the quark sector and result in no
physical phase. But many of the SM quark masses are roughly
degenerate, and many mixings in (\ref{eq:ckm}) are small, so again there is a
huge suppression. For the mixings, for example, we get a factor of
$\lambda^6 \sim 10^{-3}$.

Again, these properties do not hold for generic BSM physics, and so
the constraints thereon are strong.
\subsection{Electroweak precision tests and custodial symmetry \label{sec:ewpt}}
There is also a SM suppression in electroweak precision tests which is
not generic. To see it, consider the Higgs sector. The Higgs is a
complex $SU(2)$ doublet, and so there are four real fields. The
kinetic terms therefore have an $O(4)$ symmetry. Let us now consider
how this symmetry gets broken when we switch on the various
couplings. 

One of the miracles of group theory is that the Lie algebra of the
group $O(4)$ is the same as that of the group $SU(2) \times SU(2)$
(exercise \ex). So
the Higgs fields can be thought of as carrying 2 $SU(2)$ symmetries,
rather than the single $SU(2)_L$ of the standard model. It is usual to
call the other symmetry $SU(2)_R$, so the Higgs carries a $(2,2)$ rep (exercise \ex) of
$SU(2)_L \times SU(2)_R$.
Now, when we switch on the
$SU(2)_L$ gauge coupling $g$, we still have global symmetry $SU(2)_L$
(because the gauge symmetry includes constant gauge transformations,
which are the same as the global ones) and we still have global
symmetry $SU(2)_R$, because this factor is independent of $SU(2)_L$.
So the full $SU(2)_L \times SU(2)_R$ remains unbroken.

What is more, this $SU(2)_L \times SU(2)_R$ is also unbroken when we
switch on the
Higgs potential, because $V(H)$ is only a function of $|H|^2 = h_1^2
+h_2^2 +h_3^2 +h_4^2$, which is manifestly invariant under $O(4)$.

The Yukawa couplings do break $SU(2)_L \times SU(2)_R$,\footnote{A
  technical point: if $\lambda^u = \lambda^d$, then we can group $u^c$
and $d^c$ into an $SU(2)_R$ doublet, and $SU(2)_L
\times SU(2)_R$ is restored.} as does the
coupling to the $Z$ (which couples to the combination $T^3_L + T^3_R$).
So the correct statement is that the SM is invariant under $SU(2)_L
\times SU(2)_R$ in the limit that $\lambda^u = g^\prime =0$. 

When the Higgs gets a VEV, the $SU(2)_L \times SU(2)_R$ is broken to
the diagonal $SU(2)_V$ combination of the 2 original $SU(2)$s (exercise \ex). This
(approximate) symmetry is called `custodial $SU(2)$'. So what?
Consider the lagrangian for the gauge bosons after EWSB (the
proper way to do this is using an effective field theory, which we'll
discuss later). The remaining symmetry is just the $U(1)$ of electromagnetism
and so we should write the most general lagrangian consistent with
this. At quadratic level, in momentum space, we have
\begin{gather} \label{eq:leffw}
\mathcal{L} = \Pi_{+-} W^+W^- + \Pi_{33} W^3W^3 + \Pi_{3B} W^3 B +  \Pi_{BB} BB,
\end{gather}
where $\Pi_{ab} (p^2)$ are functions of momentum which are generated by the
currents to which the $W$ and $B$ couple: $\Pi_{ab} \sim \langle
J_a J_b \rangle $. At low energies,
we may Maclaurin expand $\Pi (p^2) = \Pi (0) + p^2  \Pi^\prime
(0) + \dots$
Consider the combination $\Pi_{+-} (0) - \Pi_{33} (0)$, which is
proportional to something called the `$T$-parameter', and which is
evidently related to the $W$- and $Z$-boson masses. Its parts are each generated
from the product of
two $W$ currents, each of which transforms as a $(3,1)$ of $SU(2)_L
\times SU(2)_R$, or a $3$ of $SU(2)_V$. The product of two 3s
decomposes as $3 \times 3 = 1 + 3 + 5$ (exercise \ex). The particular combination 
$\Pi_{+-} (0) - \Pi_{33} (0) $ is symmetric in the two indices and
traceless, so transforms as the 5 of $SU(2)_V$. But since $SU(2)_V$ is
a symmetry of the vacuuum, only singlets of $SU(2)_V$ can have
non-vanishing VEVs. This implies that $T = 0$, which in turn implies a
definite relation for, say, $\frac{m_W}{m_Z}$.

At this point, you might be wondering why I have gone through such an
arcane derivation of a result that could have easily be obtained
directly by plugging the Higgs VEV into the SM lagrangian and
computing the masses. The point is that our derivation applies not
only to the SM, but extends to any theory with $SU(2)_L \times
SU(2)_R$ symmetry that is broken to $SU(2)_V$ in the vacuum. Any such
theory will naturally come out with the right value for the measured
$T$ parameter. Conversely, BSM theories which do not feature this
symmetry
are likely to be ruled out. As examples, a model with extra Higgs scalar
states that get a VEV will have problems. For example, if we add a
Higgs triplet, then there is no approximate $SU(2)_L \times
SU(2)_R$. Even if we add only an additional Higgs doublet, we will get
in to trouble, because the theory remains  $SU(2)_L \times
SU(2)_R$ symmetric, but $SU(2)_V$ is now broken in the vacuum. One
can easily see this using the $O(4)$ language: a single complex Higgs doublet
(which can be thought of as a scalar field with values in
$\mathbb{R}^4$ breaks the group of $O(4)$ orthogonal transformations of $\mathbb{R}^4$ down to $O(3)$
(which is locally equivalent to $SU(2)_V$); a second complex Higgs
doublet will break this even further to $O(2) \simeq U(1)$, which is
just electromagnetism. As a
result, our proof above does not go through, because the vacuum is no
longer $SU(2)_V$ symmetric, and $T$ contains a singlet under the
surviving $U(1)$ (exercise \ex).
\subsection{Accidental symmetries and proton decay}
The last miracle of the SM that I want to mention is that it has accidental symmetries. These
are symmetries of the lagrangian that are not put in by {\em fiat},
but arise accidentally from the field content and {\em other} symmetry restrictions,
and the insistence on renormalizability. 

A simple example of an accidental symmetry is parity in QED.
The most general, Lorentz-invariant, renormalizable lagrangian for electromagnetism coupled to a Dirac
fermion $\Psi$ may be written as
\begin{gather}
\mathcal{L} = -\frac{1}{4} F_{\mu \nu} F^{\mu \nu} + ia F_{\mu \nu} \tilde{F}^{\mu \nu} +
i\overline{\Psi}\slashed{D} \Psi + \overline{\Psi} (m + i\gamma^5 m_5) \Psi,
\end{gather}
where both the term involving $\tilde{F}^{\mu \nu} \equiv \epsilon
^{\mu \nu \sigma \rho}F_{\sigma \rho}$ and the term involving
$\gamma^5$ na\"{\i}vely violate parity (exercise \ex). However, the former term is a
total derivative (exercise \ex) and so does not contribute to physics at
any order in perturbation theory (we will discuss important
non-perturbative contributions from such terms when we discuss the axion later
on). The latter term can be removed by a chiral rotation $\psi
\rightarrow  e^{i\alpha \gamma^5}\psi$ to leave a parity-invariant
theory with fermion mass $\sqrt{m^2 + m_5^2}$. So we find that the
lagrangian is invariant under parity, even though we did not require
this in the first place. The same is true of charge conjugation
symmetry. Note that if we had not insisted on renormalizability, we
could write dimension-six terms like $\overline{\Psi} \gamma^\mu
\gamma^5 \Psi \overline{\Psi} \gamma_\mu \Psi$, which do violate
parity (exercise \ex: show this violates $P$). We will explore these when we discuss effective field theories
later on.

As we already alluded to above, the SM lagrangian is accidentally
invariant under a $U(1)_B$ baryon number symmetry (an overall rephasing
of all quarks)
and three $U(1)$ lepton number symmetries, corresponding to individual
rephasings of the three different lepton families (which contains an
overall lepton number symmetry $U(1)_L$ as the diagonal subgroup).
Either $U(1)_B$ or $U(1)_L$  symmetry, together with Lorentz invariance, prevents the proton from
decaying. Indeed, a putative final state must (by Lorentz invariance,
which implies the fermion number is conserved mod 2) contain an odd number of fermions lighter than the proton. The only such states
carry lepton number but not baryon number, whereas the proton carries
baryon number but not lepton number.

Again, once we allow higher dimension operators, we will find that
lepton and baryon number are violated (by operators of dimension five
or six, respectively), meaning that the proton can decay. Similarly, generic theories of physics BSM will
violate them and hence will be subject to strong constraints.

(Exercise \ex: consider just the Higgs sector coupled to the
$W$-boson. Show that $SU(2)_L \times SU(2)_R$ is an accidental
symmetry, and find a dimension-six operator that violates it.)

\subsection{What isn't explained}
We have now seen that the SM does a wonderful job of explaining a lot
of data, and it does so by means of a delicate structure that is not
preserved by generic BSM theories. This means that it is hard to write
down BSM models that are consistent with all the data. 

But we are impelled to write down models by the fact that there are, by now, also plenty of
data that the SM patently cannot describe. These include:
\begin{enumerate}[(i)]
\item {\em neutrino masses and mixings}\hfill \\
No term in (\ref{eq:lyuk}) yields these.
\item {\em the presence of non-baryonic, cold dark matter}\hfill \\
Dark matter is neutral, colourless, non-baryonic, and massive. The only such particles in
the SM are neutrinos, but these are too light, making instead warm dark
matter.
\item {\em the presence of scale-invariant, Gaussian, and apparently acausal density perturba-
tions, consistent with a period of inflation at early times}\hfill \\
\item {\em the observed abundance of matter over anti-matter}\hfill \\
I have yet to meet my {\em alter ego} and indeed there would appear to
be a general predominance of matter over anti-matter in the
Universe. Note, moreover, that inflation would destroy any asymmetry
imposed as an initial condition. 
\end{enumerate}
All of these have been established beyond reasonable doubt, in many
cases overwhelmingly so. In addition, there are many unexplained
features of nature that lead us to believe that there must be physics BSM:
\begin{enumerate}[(i)]
\item {\em the inability to describe physics at planckian scales}\hfill \\
Contrary to what you might read in the New Scientist, general
relativity makes perfect sense as 
a theory of quantum gravity 
up to planckian scales (as an effective field theory, to be described below), but beyond that we need a theory
of quantum gravity, such as string theory.
\item {\em the hierarchy between the observed cosmological constant and
  other scales} \hfill \\
The measured energy density associated with the accelerated expansion
of the Universe is $(10^{-3} \mathrm{eV})^4$, but receives
contributions of size $(\mathrm{GeV})^4, (\mathrm{TeV})^4,$ {\em \&c.}
from QCD, weak scale physics, {\em \&c}. Such a small value appears
necessary to support intelligent life (and you and me), but how is it achieved?
\item {\em the hierarchy between the weak and other presumed scales}
  \hfill \\
As above, but now the question is how to get a TeV from, {\em e.g.}
the Planck scale. Here, it is less clear that such a small value is a
{\em sine qua non} for us to be here worrying about it in the first place.
\item {\em the comparable values of matter, radiation, and vacuum
    energy densities today} \hfill \\
These 3 scale in vastly different ways during the Universe's evolution, so why are they
roughly the same today? 
\item {\em the structure in fermion masses and mixings}\hfill \\
There is a hierarchical structure in these (as described, {\em e.\
  g.}, in \cite{Giudice:2011ak}), but in the SM they are
just free parameters. So why do they exhibit structure?
\item {\em the smallness of measured electric dipole moments}\hfill \\
These violate $CP$, and are at least $10^{-10}$ smaller than a
na\"{\i}ve guess of $O(1)$ for the corresponding parameter, which is
an angle $\in [0,2\pi]$; why?
No anthropic argument is known, by the way.
\item {\em the comparable size of the 3 gauge couplings}\hfill \\
These are all $O(1)$ and different, but not {\em so} different; why?
\item {\em the quantization of electric charges}\hfill \\
The SM contains the gauge group $U(1)_Y$, for which any charge is
allowed.
Why do we find integer multiples of $\frac{1}{3}$ for the electric
charge, rather than, {\em e.\ g.} $\sqrt{2}$ or $\pi$?
\item {\em the number of fermion families}\hfill \\
Why 3? As Rabi said about the muon, `Who ordered that?'
\item {\em the number of spacetime dimensions}\hfill \\
Why 4? Why 3+1 for that matter?
\end{enumerate}

An explanation (together with an experimental confirmation, of course) of any one of these would surely merit a Nobel
prize, not least because they are such deep issues, but also because it seems
so hard to extend the SM in such a way as to furnish a solution,
without contravening some other experimental test.

I will certainly not provide definitive answers to any of them in
these lectures, or even address any of them in detail.
Rather, I aim to introduce you to one or two of the ideas that have
been proposed, in order to give you a flavour of the way that the game
is played. But before I do that, I introduce a framework that enables
us to discuss many of the issues in BSM physics in a generic,
model-independent way. This requires us to learn a little about
effective field theory. 
\section{Beyond the SM - Effective field theory}
Once we go BSM, it seems like an infinity of possibilities opens up --
we could write down any lagrangian we like. Fortunately, we have a
good starting point, since we know that we must reproduce the SM in
some limit. 

Even better, we can make things very concrete by making one assumption. Let
us suppose that any new physics is rather heavy. This is indicated experimentally
by the fact
that observed deviations from the SM are small, but it is not the only
possibility. New physics could instead be very light, but also very weakly
coupled to us. We'll see an example of this later on, when we mention
large extra dimensions. 

With this asumption in hand, we can analyse physics BSM in a completely
general way using methods of effective field theory (EFT).

\subsection{Effective field theory}
To motivate the EFT idea (about which we shall be scandalously brief;
for more thorough treatments, see \cite{Polchinski:1992ed,Manohar:1996cq,Rothstein:2003mp,Kaplan:2005es}), suppose we start with the renormalizable SM,
and consider only energies and momenta well below the weak scale,
$\sim 10^2$ GeV. We can never produce $W$, $Z$, or $h$ bosons on-shell
and so we can simply do the path integral with respect to these
fields (we `integrate them out', to use the vernacular). At tree-level,
this just corresponds to replacing the fields using their classical
equations of motion, and expanding $\frac{-1}{q^2-m_W^2} =
\frac{1}{m_W^2} + \frac{q^2}{m_W^4} + \dots$. It is already clear that
our expansion breaks down for momenta comparable to $m_W$, so that the
theory is naturally equipped with a cut-off scale. We will be left
with a path integral for the light fields, but with a complicated
lagrangian that is non-local in space and time. But since we are only interested in low energies
and momenta, we can expand in powers of the spacetime derivatives (and
the fields)
to obtain an infinite series of {\em local} lagrangian operators, which become
less and less important as we go down in (energy-)momentum. We can use this
theory to make predictions to any desired order in the momentum
expansion simply by retaining sufficiently many terms. Many of these
operators are, of course, non-renormalizable, but this is unimportant
since the theory is naturally equipped with a UV cut-off $\sim 10^2$
GeV, beyond which we know that it breaks down. So we can simply cut
off our loop integrals there and never have to worry about
divergences. We call such a theory
an EFT. The rules for making an EFT are exactly the same as those for
making a QFT, except that we no longer insist on
renormalizability. Instead, we specify the fields and the symmetries,
write down all the possible operators, and accept that the theory will
come equipped with a cut-off, $\Lambda$, beyond which the expansion
breaks down.

So let us, in this way, now imagine that the SM itself is really just an
effective, low-energy description of some more complete BSM theory. 
Thus, the fields and the (gauge) symmetries of the theory are exactly the same
as in the SM, but we no longer insist on renormalizability. For
operators up to dimension 4, we simply recover the SM. But at
dimensions higher than 4, we obtain new operators, with new physical
effects. As a striking example of these, we expect that the accidental
baryon and lepton number symmetries of the SM will be violated at
some order in the expansion, and protons will decay.\footnote{Let us
  hope that we can finish the lecture before they do so!} 

We don't know what the BSM theory actually is yet, and so when we
write down the EFT, we should allow the coefficients of the
operators in the expansion to be arbitrary. Since there are infinitely
many such operators, and infinitely many coefficients, one might worry
that this means that predictivity is lost -- it seems that we need to
make infinitely many measurements (to fix all the coefficients) before
we can make predictions. But this is not true, once we truncate the
theory at a given order in the operator/momentum expansion: at any given finite
order,
the number of coefficients is finite and so we can eventually make
predictions for observables, with a finite precision that is fixed by the
truncation of the momentum expansion. 

Before we discuss the specific operators that arise in the SM and
their physical effects, it is useful to make a few technical points
about EFTs.

The first point is that, while we don't know the actual values of the
coefficients, we can estimate their size using dimensional
analysis, since we expect the expansion to break down at energies of
order the cut-off, $\Lambda$. So the natural size of coefficients is typically
just an $O(1)$ number in units of $\Lambda$.

The second point is that the operators of a given dimension form a
vector space, and so it is useful to choose a basis for these. 
This is not so straightforward as it sounds (and indeed, there are still
disputes about it in the literature from time to time), because of
equivalences between operators. In particular, any two operators that
are equal up to a total divergence may be considered equal (since they
give the same contribution at any order in perturbation theory), as may
operators that differ by terms that vanish when the equations of
motion hold, because such pieces can be removed by a
field redefinition in the path integral (see, {\em e.\ g.}, \cite{Arzt:1993gz}).

The third point concerns loop effects. We previously argued
that operators of higher and higher dimension give smaller and smaller
contributions to low energy processes. This is true for tree-level
diagrams, but it is not obviously true when we insert these operators
into loops, and integrate over all loop momenta up to the cut-off
$\Lambda$.
Apparently then, all higher-dimension operators are unsuppressed in loop
diagrams.
This looks like a disaster because it appears that we cannot truncate
the expansion when we include loops. But we are saved by the fact that
the only effect of such diagrams, once we expand them in powers of the
external momenta, is to generate corrections to lower
dimensional operators. So all these loops do is to correct the
coefficients of other operators. This suggests that there should exist
a regularization scheme in which these corrections are already taken
into account, which will be far more convenient than regulating with a
hard cut-off. The `right' scheme is dimensional
regularization, because then $\Lambda$s don't appear in the numerators of the loop amplitudes.
 The upshot is that if you
ever have to compute a loop diagram in EFT, you should do it using
dimensional regularization.\footnote{For an example of
  what happens if you don't, see \cite{Gripaios:2013lea}.}

A fourth point: if non-renormalizable EFTs make sense, why did we ever insist
on renormalizability ? Well, a renormalizable theory can now be
thought of 
as a special case of a non-renormalizable theory, in which we take the
cut-off to be very large. Then, the operators with dimension greater
than 4 become completely negligible (hence they are known as
`irrelevant' operators in the jargon), operators with dimension =4
stay the same (hence they are `marginal'), and operators with
dimension <4 dominate (and are called `relevant').  So if we observe
physics that is well described by a renormalizable theory, we should
conclude that the cut-off (at which new physics presumably appears) is
rather far away.

Finally, we can see that there is a big problem with relevant operators in
EFTs. Consider for example, the mass of a scalar field in an EFT with
cut-off $\Lambda$. By the arguments above, our estimate for the size
of the mass (which is just a dimensionful operator coefficient) is $m
\sim \Lambda$. But then the EFT is not of much use,
because the particle can never be produced in the regime of validity
of the EFT. So, unless there is some dynamical mechanism or tuning
that makes the mass rather smaller than our estimate, the EFT does not
make sense. The same is true for any relevant operator, and this leads
to the hierarchy problems of the SM that we discuss in more detail below.

Now we know vaguely what an EFT is, and how to do computations using
one, we can discuss the extension of the SM to an EFT and the
operators that arise, starting with the most relevant ones. 
\subsection{$D=0$: the cosmological constant}
We have avoided mentioning it up to now, but clearly a constant term
(which has dimension 0) is
consistent with the symmetries of the SM. It has no effect until the
SM is coupled to gravity, whereupon it causes the Universe to
accelerate. On the one hand, this looks like good news, because the
Universe is observed to accelerate. On the other hand, this is bad
news because our estimate of the size of this operator coefficient
(the operator is 1) is $\Lambda^4$, while the observed energy density is around $(10^{-3}~
\mathrm{eV})^4$. But the cut-off of the SM had better not be $10^{-3}~
\mathrm{eV}$, because if it were then we could certainly not use it to
make predictions at LHC energies of several TeV. So either dynamics or
a tuning makes the constant small. If we consider the Planck scale to
be to be a real physical cut-off, then we need to tune at the level of 1 part in $10^{120}$. It
is fair to say, that despite $O(10^{120})$ papers having been written
on the subject, no satisfactory
dynamical solution has been suggested hitherto. An alternative is to argue that
we live in a multiverse in which the constant takes many different
values in different corners, and we happen to live in one which is
conducive to life. Indeed, it has been argued \cite{Weinberg:1988cp} that if the constant
were much larger and positive, structure could never form, while if it
were too large and negative, the Universe would re-collapse before
life could appear. The flavour-of-the-month as regards how the
multiverse itself arises is by a process of eternal inflation in
string theory.
\subsection{$D=2$: the Higgs mass parameter}
The only other relevant operator in the SM is the Higgs mass
parameter, which sets the weak scale. As above, the natural size for this is
$\Lambda$. But we measure $v \sim 10^2$ GeV, leaving us with 2
options:
either the natural cut-off of the SM is not far above the weak scale
(in which case we can hope to see evidence for this, in the form of new
physics, at the LHC) or the cut-off is much larger, and the weak scale
is tuned, perhaps once again by anthropics. 
\subsection{$D=4$: marginal operators}
We have discussed these already in the context of the renormalizable
SM, and there is nothing to add here. 
\subsection{$D=5$: neutrino masses and mixings}
Now things get more interesting. There is precisely one operator at
$D=5$, namely $\frac{\lambda^{ll}}{\Lambda} (lH)^2$, where
$\lambda^{ll}$ is a dimensionless $3\times 3$
matrix in flavour space. Note that this operator violates the
individual and total lepton numbers; moreover, it gives masses to
neutrinos after EWSB (exercise \ex), just as we observe. So, one might argue that it is no surprise that neutrino masses
have been observed, since they represent the {\em leading} deviation from
the SM, in terms of the operator expansion.
Given the observed $
10^{-3} ~\mathrm{eV}^2$ mass-squared
differences of the neutrinos, we estimate $\Lambda \sim 10^{14}$
GeV. Thus, one could argue
that while neutrino masses are undeniably, as one so often hears, evidence for physics BSM,
they are also evidence that the SM is valid up to energy scales that
are way, way beyond the reach of conceivable future colliders.

Even so, it is worthwhile to consider what theory might replace the
EFT at $\Lambda$ to give a UV completion, extending the regime of
validity. One extremely simple possibility is to add to
the SM a new fermion, $\nu^c$, that is a singlet under $SU(3)\times SU(2) \times
U(1)$. In fact we need at least 2 of these to generate the two
observed neutrino mass-squared differences, and it seems plausible
that there are 3 -- one for each SM family.

We may then replace the $D=5$ operator with the renormalizable Yukawa
term $\lambda^\nu l H^c \nu^c$ (which is a Dirac mass term for
neutrinos after EWSB), along with the Majorana mass term $m^\nu \nu^c
\nu^c$. This leads to the so-called `see-saw' mechanism, about which
you may have heard. (Exercise \ex: Count the mixing angles and phases in
the lepton sector, in the presence of either or both of these terms.)
(Exercise \ex: How is $\lambda^{ll}$ related to $\lambda^\nu$ and $m^\nu$?)

Finally, note that the other neutrino mass eigenstates in this
renormalizable model need not be heavy. Indeed, they could be much
lighter, but very weakly coupled to SM states.
\subsection{$D=6$: trouble at t'mill}
Once we get to $D=6$, a whole slew of operators appear. These include
operators that violate baryon and lepton number, such as $\frac{qqql}{\Lambda^2}
$ and $\frac{u^c u^c d^c e^c}{\Lambda^2}$
(exercise \ex: check these are invariants), and which cause the
proton to decay via $p \rightarrow e^+ \pi^0$. We can estimate a lower bound on $\Lambda$ from the experimental bounds on the proton
lifetime, $\tau_p > 10^{33}$ yr, as follows. The decay rate (which comes from the amplitude
squared) is proportional to $\frac{1}{\Lambda^4}$ and the remaining
dimensions must be supplied by phase space, giving a factor of
$m_p^5$. Plugging in the numbers (exercise \ex), we get $\Lambda > 10^{15}$
GeV. Again, the implication is that new physics either respects baryon
or lepton number, or is a long way away.

There are also operators that give corrections to flavour-changing
processes that are highly suppressed in the SM, for reasons already
given. As an example, the operator $(s^c d) (d^c s)/\Lambda^2$ contributes to Kaon mixing
and measurements of $\Delta m_K$ and $\epsilon_K$ yield a bound of
$\Lambda > 10^5$ TeV. 
\section{Grand Unification}
We now turn to discuss in more detail some of the hints for physics
BSM. Perhaps the most compelling of these is the apparent unification
of gauge couplings.

As you know, 
one consequence of renormalization in QFT is
that the parameters of the theory must be interpreted as being
dependent on the scale at which the theory is probed. The QCD
coupling and $g$ get smaller as the energy scale goes up,
while $g^\prime$ gets
larger. Remarkably, if one extrapolates far enough, one finds that all
three couplings are nearly\footnote{Nearly enough to be impressive, but not quite. The
  discrepancy is resolved in the MSSM, however.}
equal\footnote{At the moment, this is an trivial statement: the
  normalization of $g^\prime$ is arbitrary and can always be chosen to
  make all three couplings meet at the same point. But we will soon be
able to give real meaning to it.} at a very high scale, {\em c.} $10^{15}$ GeV. Could it be that,
just as electromagnetism and the weak force become the unified
electroweak force at
the 100 GeV scale, all three forces become unified at $10^{15}$ GeV? 

The fact that the couplings seem to become equal is a hint that we
could try to make all three groups in $SU(3) \times SU(2) \times U(1)$
subgroups of one big group, with a single coupling constant. We need a
group with rank at least 4 (exercise \ex: why?), where the rank is the maximal number of
commuting generators in a basis for the Lie algebra.
The group $SU(5)$ is an obvious contender (exercise \ex: prove that
$SU(N)$ has rank $N-1$). How does $SU(3) \times SU(2) \times U(1)$ fit into
$SU(5)$? Consider $SU(5)$ in terms of its defining rep: 5 $\times$
5 unitary matrices with unit determinant acting on 5-d vectors. We can
get an $SU(3)$ subgroup by considering the upper-left 3 $\times$ 3 block and
we can get an independent $SU(2)$ subgroup from the lower right 2
$\times$ 2
block. There is one more Hermitian, traceless generator that is
orthogonal to the generators of these two subgroups: it is
$T = \sqrt{\frac{3}{5}}\mathrm{diag} (-\frac{1}{3}, -\frac{1}{3}, -\frac{1}{3}, \frac{1}{2},
\frac{1}{2})$, in the usual normalization. Our goal will be to try
to identify this with the hypercharge $U(1)_Y$ in the SM. To do so, we
first have to work out how the SM fermions fit into reps of
$SU(5)$. 

Before going further, let's do a bit of basic $SU(N)$ representation
theory. The defining, or {\em fundamental}, representation is an
$N$-dimensional vector, $\alpha^i$,
acted on by $N \times N$ matrices. We can write the action as
$\alpha^i \rightarrow U^i_j \alpha^j$, with the indices $i,j$
enumerating the $N$ components. Given this rep, we can immediately
find another (at least for $N>2$) by taking the complex
conjugate. (Exercise \ex: prove that the 2-d irrep of $SU(2)$ is pseudo-real.) This is called the
{\em antifundamental} rep. It is convenient to
denote an object which transforms according to the antifundamental
with a downstairs index, $\beta_i$. Why? The conjugate of $\alpha^i
\rightarrow U^i_j \alpha^j$ is $\alpha^{*i} \rightarrow U^{*i}_j
\alpha^{*j} = U^{\dagger j}_{i}\alpha^{*j}$. So if we define things
that transform according to the conjugate with a downstairs index, we
can write $\beta_i \rightarrow U^{\dagger j}_{i} \beta_j$.
The beauty of this is that $\alpha^i \beta_i \rightarrow \alpha^j
U_j^i U^{\dagger k
}_i \beta_k = \alpha^j
\delta_j^k \beta_k = \alpha^k \beta_k $, where we used $UU^\dagger
=1$. Thus when we contract an upstairs index with a downstairs index, we get
a singlet. This is, of course, much like what happens with $\mu$,
$\alpha$, and $\dot{\alpha}$ indices
for Lorentz transformations. Note that the Kronecker delta,
$\delta_j^k$, naturally has one up index and one down and it transforms as
$\delta^i_l \rightarrow U^i_k \delta_j^k U^{\dagger j}_l$.
But $UU^\dagger
=1 \implies \delta_i^l \rightarrow \delta_i^l$ and so we call
$\delta_i^l$ an {\em invariant tensor} of $SU(N)$. Note, furthermore,
that there is a second invariant tensor, namely $\epsilon_{ijk\dots}$
(or $\epsilon^{ijk\dots}$),
the totally antisymmetric tensor with $N$ indices. Its invariance
follows from the relation $\mathrm{det} \; U =1$.

These two invariant tensors allow us to find all the irreps
$SU(N)$ from (tensor) products of fundamental and antifundamental
irreps. The key observation is that tensors which are
symmetric or antisymmetric in their indices remain symmetric or
antisymmetric under the group action (exercise \ex), so cannot transform into one
another. So to reduce a generic product rep
into irreps, one can start by symmetrizing or antisymmetrizing the
indices. This doesn't complete the process, because one can also
contract indices using either of the invariant tensors, which also produces
objects which only transform among themselves (exercise \ex). 

Let's see how it works for some simple examples. Start with $SU(2)$, which is locally equivalent to $SO(3)$ and
whose representation theory is known to the man on the
Clapham omnibus as `addition of angular
momenta in quantum mechanics'. The fundamental rep is a 2-vector
(a.k.a. spin-half); call it $\alpha^j$. Via the invariant tensor
$\epsilon_{ij}$ this can also be thought of as an object with a downstairs
index, {\em viz.} $\epsilon_{ij} \alpha^j$, reflecting the fact that the doublet
and anti-doublet are {\em equivalent} representations. So
all tensors can be thought of as having indices upstairs, and it
remains only to symmetrize (or antisymmetrize). Take the product of two
doublets for example. We decompose $\alpha^i \beta^j =
\frac{1}{2}(\alpha^{(i} \beta^{j)} +\alpha^{[i} \beta^{j]}  )$, where
we have explicitly (anti)symmetrized the indices. The
symmetric object is a triplet irrep (it has $(11)$, $(22)$, and $(12)$
components), while the antisymmetric object is a singlet (having only
a $[12]$ component). We write this decomposition as $2 \times 2 = 3 +
1$, where we label the irreps by their dimensions.\footnote{One has to
  be careful doing this: $SO(5)$ for example, has 2, inequivalent
  30-dimensional irreps \cite{Gripaios:2014pqa}.}

The representation theory of $SU(3)$ is not much harder. The
fundamental is a triplet and the anti-triplet is
inequivalent.\footnote{It is inequivalent, because we cannot convert
  one to the other using $\epsilon_{ij}$, which has been replaced by $\epsilon_{ijk}$.} The
product of two triplets contains a symmetric sextuplet and an
antisymmetric part containing three states. We can use the invariant
tensor $\epsilon_{ijk}$ to write the latter as
$\epsilon_{ijk}\alpha^{[i} \beta^{j]}$, meaning that it is equivalent
  to an object with one index downstairs, {\em viz.} an
  anti-triplet. Thus the decomposition is $3 \times 3 = 6 + \overline{3}$. On the
  other hand, we cannot symmetrize the product of a 3 and a
  $\overline{3}$, because the indices are of different type. The
  only thing we can do is to separate out a singlet obtained by
  contracting the two indices with the invariant tensor
  $\delta^i_j$. Thus $3 \times \overline{3} = 8 + 1$ (exercise \ex:
  decompose $\alpha^i \beta_j$ explicitly). The 8 is
  the adjoint rep. Again, the man on the Clapham omnibus calls this `the eightfold way'.

For $SU(5)$, things are much the same. The only reps we shall need are
the smallest ones, namely the (anti)fundamental 5($\overline{5}$) and
the 10 which is obtained from the antisymmetric product of two 5s.

Now let's get back to grand unified theories. We'll try to do the dumbest thing imaginable which is to try to fit
some of the SM particles into the fundamental five-dimensional
representation of $SU(5)$. I hope you can see that, under $SU(5)
\rightarrow SU(3) \times SU(2) \times U(1)$, this breaks up
into a piece (the first three entries of the vector) that transform
like the fundamental (triplet) rep of $SU(3)$ and the singlet of $SU(2)$ and a
piece (the last two entries of the vector) which is a singlet of
$SU(3)$ and a doublet of $SU(2)$. For this to work the last two entries would have to
correspond to $l$ (since this is the only SM multiplet which is a
singlet of $SU(3)$ and a doublet of $SU(2)$), in which case the hypercharge must be fixed to be
$Y = -\sqrt{\frac{5}{3}} T$. Then the hypercharge of the first three
entries is $+\frac{1}{3}$. This is just what we need for $d^c$,
except that  $d^c$ is a colour anti-triplet rather than a
triplet. But we can fix it up by instead identifying $Y =
+\sqrt{\frac{5}{3}} T$ and then identifying $(d^c , l)$
with the {\em anti-fundamental} rep of $SU(5)$.

What about the other SM fermions? The next smallest rep of $SU(5)$ is
ten dimensional. It can be formed by taking the product of two
fundamentals and then keeping only the antisymmetric part of the
product. But since we now know that under $SU(5) \rightarrow 
SU(3) \times SU(2) \times U(1)$, $5 \rightarrow (3,1,-\frac{1}{3}) +
(1,2,+\frac{1}{2})$, you can immediately deduce (exercise \ex) that $10 \rightarrow
(3,2,+\frac{1}{6}) + (\overline{3},1,-\frac{2}{3})+ (1,1,+1)$.
These are precisely $q, u^c,$ and $e^c$. 

That things fit in this way is nothing short of miraculous. Let's now
justify our statement about the couplings meeting at the high
scale. The $SU(5)$ covariant derivative is
\begin{gather}
D_\mu = \partial_\mu + i g_\mathrm{GUT} A_\mu
\supset ig_\mathrm{GUT} \left(W^3_\mu T^3 + i \sqrt{\frac{3}{5}} Y
  B_\mu \right),
\end{gather} 
so unification predicts that $\tan \theta_W = \frac{g}{g^\prime} =
\sqrt{\frac{3}{5}} \implies \sin^2 \theta_W = \frac{3}{8}$. {\em This}
is the relation which is observed to hold good (very nearly) at the
unification scale, $\Lambda \sim 10^{15} ~\mathrm{GeV}$. 
This scale is very high, which is bad news for testing
unification. But it is just as well, since it is clear that baryon and
lepton number cannot be symmetries of this model (exercise \ex: why
not?). So protons decay, and indeed $\Lambda$ is right around the
bound therefrom. When GUTs were first put forward, this led to high
hopes that protons would be observed to decay, and many experiments were
carried out. So far, no dice.

There is another GUT which is based on the group $SO(10)$. This is
perhaps even more remarkable, in that the fifteen states of a single
SM generation fit into a 16 dimensional rep (it is in fact a spinor)
of $SO(10)$. You might be thinking that this doesn't look so good, but
--- wait for it --- the sixteenth state is a SM gauge singlet and plays
the r\^{o}le of $\nu^c$. 

At this point, it is worthwhile to pause and to reflect on just how
much has been explained. It is already a surprise in the SM that the 3
gauge couplings are remotely comparable in size at low energy. After
all, we know only that they should be at most $O(1)$ for
perturbativity; tiny values for one or more of them would seem to be
fine.
Unification explains not only why they are comparable, but also
gives a precise prediction for them which seems, very nearly, to
hold. Even more, the predicted scale of unification is exactly where
it needs to be: somewhat below the (ultimate?) Planck scale, but just
above the bound implied by the proton lifetime. Finally, the SM
multiplets fit precisely into the smallest multiplets of the GUT
groups, with no missing or extra states, and charge quantization is
explained, because hypercharge lies in an underlying non-Abelian
group. How can this not be correct, you might wonder?

(Exercise \ex: show that a scalar particle in an adjoint of $SU(5)$ can
achieve the required breaking of $SU(5) \rightarrow SU(3) \times SU(2)
\times U(1)$.)

\section{Approaches to the hierarchy problem}
In a nutshell, the electroweak hierarchy problem is to explain how the
weak scale, $\sim 100 ~\mathrm{GeV}$, emerges from a more fundamental theory
with (presumably) much higher scales, like $m_{GUT}$, $m_P$ or even
the indicated neutrino mass scale $\sim 10^{14}$ GeV. This would be a
problem even in a classical theory. In QFT, it is even
worse, because quantum corrections will typically generate all
operators at low energy that are not forbidden by
symmetries, with size set by the cut-off.\footnote{This is sometimes called Gell-Mann's totalitarian
  principle: everything which is not forbidden is compulsory.}
To see the problem, consider a lagrangian with two scalar fields,
$\phi$ and $\Phi$, with masses $m$ and $M (\gg m)$ respectively. 
Loops of $\Phi$ fields will generate corrections of order $M$ to
$m$. So to keep the mass of $\phi$ light, we need to very carefully
tune these corrections against an $O(M)$ bare mass for $\phi$, in
order for a small mass $m$ to emerge at low energies. More generally,
in any theory with a heavy mass scale, we expect quantum corrections
to lift light masses up to the heavy scale, unless some delicate
mechanism prevents it. 

It is worth pointing out that we do know of mechanisms by which
particles can remain light in the presence of heavier scales. For
example, because of chiral symmetry, quantum corrections to a light Dirac fermion
mass $m$ must be proportional to $m$. 

To see this, let us suppose $m$ is actually a field. We can then assign it
a charge such that the chiral symmetry is restored. Explicitly, $(q,
q^c) \rightarrow e^{i\alpha} (q,q^c)$ and $m \rightarrow e^{-2i\alpha}
m$.

Now consider the quantum corrections to $m$, $\delta m$. $\delta m$
must transform in the same way as $m$ under the chiral symmetry, and
so covariance implies that the expression for $\delta m$ (which cannot
involve the fields $q$ and $q^c$) must be of the form
\begin{gather}
\delta m = m f(|m|^2),
\end{gather}
where $f$ is regular at the origin. This tells us immediately that $\delta m$ is small when $m$ is small. 

This type of argument is a powerful one, of very general
applicability, and so we pause to examine it further. To make it, we
take some parameter of the theory and observe that the theory has some enhanced
symmetry (here chiral symmetry) if we allow that parameter to transform in a
certain way. We can use this to work out how the
parameter must appear in various expressions, by insisting that the
symmetry is respected. One way to think about this is to imagine that
we are pretending that the parameter is just an additional field in
the theory, and so we call it a {\em spurionic} field, or just a {\em
  spurion}. We'll use these ideas again later on.

In the case at hand, the argument explains how the electron mass, for
example, can remain so small in the presence of other heavier
scales. But it doesn't explain why the electron mass is so small in
the first place!

There is also a mechanism by which scalar particles can be
light. Indeed, we know that Goldstone bosons are massless, because of
symmetry. If we have an approximate symmetry of this type, then we end
up with a naturally light {\em pseudo}-Goldstone boson.
\subsection{SUSY}
Another way to make a scalar field (like the Higgs) light is to tie
its mass to the mass of a fermion (which, as we have just argued, can
be naturally light). Miraculously, this can be achieved by enlarging
the Poincar\'{e} invariance of spacetime to a supersymmetry, with
extra symmetry generators that take bosons into fermions and {\em vice
  versa}. This is covered in other lectures, and I make only 2 remarks
here. The first is that supersymmetry is necessarily broken in Nature,
and one still needs to explain how the supersymmetry breaking scale
itself is generated in a natural way. This is not a huge problem,
however, in that supersymmetry breaking (see, {\em e.\ g.} \cite{Luty:2005sn}) can (and probably does) take
place in a {\em hidden} sector, and there are known ways to do it. The
second remark is that supersymmetry only forces the masses of fermions
and bosons to be the same: they do not necessarily have to be
light. In particular, in the minimal supersymmetric extension of the
SM (the MSSM), there is a mass term for the Higgs bosons, and one
needs a mechanism to make this mass small. This is called the $\mu$
problem; again, ways to do it are known.
\subsection{Large Extra Dimensions}
Another way to solve the hierarchy problem is to suppose that there
aren't any high scales. In particular, it may be that the Planck scale
of gravity emerges somehow from much lower energy scales. One way to
do this is to suppose that there are really $n >3$ space dimensions,
with the extra $n-3$ dimensions having size $R$. If so, the effective
$3+1$ dimensional Planck constant is given in terms of the fundamental
scale $m$ of $n+1$-d gravity by $m_P^2 = m^{n-1} R^{n-3}$ (exercise \ex).
If $m \sim$ TeV, the hierarchy problem goes away, but we find that the
extra dimensions have radius $R \sim 10^{13}$ m for $n=1$ (even most
theorists would probably have noticed this!), $R \sim 10^{-3}$ m for
$n=2$ (the realization that gravity had only been tested down to
comparable distances sparked a frenzy of tests in the last decade, and
we have now got down to about 10 $\mu$m) and $R \sim 10^{-8}$ m for
$n=3$ (good luck testing this). Again, 2 remarks are in order. The
first is that we again have not actually solved the hierarchy problem; we have turned it into the question of why the extra dimensions are
so large (again, there are ideas for how to achieve this). The second
is that such large extra dimensions give another example of physics
BSM that cannot be described by an EFT. Why not? Theories with extra
dimensions have towers of Kaluza-Klein excitations (like the vibrating
modes of a guitar string), with masses in units of $1/R$. So these new
states are extremely light, for the relevant values of $R$. For more
details, see, {\em e.\ g.}, \cite{Rattazzi:2003ea,Gherghetta:2006ha}.
\subsection{Composite Higgs}
In considering the problem of the weak scale hierarchy, it is
worthwhile to note that there is a large hierarchy in physics that is well understood. The hierarchy in question is that between
the mass of the proton, $m_p \sim$ GeV, and higher scales. It is
explained by the logarithmic running of the QCD coupling
constant. This starts off small at high energies and slowly increases
as we go down in energy. Eventually, it becomes large enough that QCD
confines, creating a low physical scale by dimensional transmutation.

The example of QCD is relevant to our discussion for another
reason. As well as confinement, the strong coupling regime of QCD
leads to the breaking of the approximate chiral symmetries acting on
light quarks. This breaking of chiral symmetry would have led to a
perfectly acceptable and natural breaking of electroweak symmetry in
the SM, even if there had been no Higgs at all!

To see this in more detail, let us consider the SM without a Higgs,
and with only up and down quarks, for simplicity. The global symmetry of the quark sector (in
the absence of EW interactions) is $SU(2)_L \times SU(2)_R \times
U(1)_B$. When we switch on the EW interactions, we gauge an $SU(2)_L
\times U(1)_Y$ subgroup of this, where $Y = T_R^3 + B/2$. Now, when
the QCD coupling becomes strong, $SU(2)_L \times SU(2)_R \times
U(1)_B$ gets spontaneously broken to $SU(2)_V \times U(1)_B$,
resulting in 3 massless Goldstone bosons (the 3 pions of QCD). But
these pions get eaten in the usual way by the EW gauge fields, since the
EW gauge symmetry is broken from $SU(2)_L
\times U(1)_Y$ to $U(1)_Q$, where $Q = T_L^3 +T_R^3 + B/2 = T_L^3 +
Y$.

So the pattern of gauge symmetry breaking is exactly what we observe
in Nature, and moreover, since the theory has the custodial symmetry
we mentioned earlier, we are guaranteed to get the right ratio of $W$
and $Z$ boson masses. 

There is one small problem however, which is that the absolute masses
of the $W$ and $Z$ are way too small! How small? Well, the $W$, say,
gets its mass from diagrams mixing it with the pions, through a vertex
coming from
the matrix element between the weak current and the
pion
\begin{gather}
\langle 0 |J_\mu^+ | \pi^-(p)\rangle \equiv i \frac{f_{\pi}}{\sqrt{2}} p_\mu
\end{gather}
and yields (exercise \ex: show this, without worrying about factors of 2) $m_W = \frac{g f_{\pi}}{2} = 29$ MeV.

Evidently, this is a disaster, but it is easy to get from here to a
more viable model, as follows. We simply imagine that there is another
gauge group, called {\em technicolour}, which becomes strongly coupled by slow running of its
coupling constant, but at a TeV rather than a GeV.\footnote{Note that
  the technicolour gauge group doesn't have to be $SU(3)$: the only
  desideratum is that it have the same pattern of chiral symmetry
  breaking as QCD.}

Technicolour was a fantastic idea, but it doesn't work, not least
because, contrary to recent experimental evidence, it does not predict
a Higgs
boson!
But we were already fairly sure that Technicolour was wrong {\em before} the
Higgs discovery, because of problems with other electroweak precision
observables (and because of problems with flavour physics, which we
discuss shortly). 
Particularly problematic was the so-called `S-parameter', which in the
EFT language of (\ref{eq:leffw}), can be written as
$\Pi^\prime_{3B}$, and which is believed to be too
large in technicolour models.\footnote{I say `believed to be', because we cannot actually
  calculate these quantities in a strongly coupled gauge theory,
  except for gauge groups of large rank $N$, in which case they are of
  $O(N)$. Our best estimate for their size near $N \sim 1$ is
  therefore $O(1)$, compared to measured values of $O(0.3)$, the
  latter agreeing well with the SM.} Now we know that $\Pi_{3B}$ is non-zero in the
vacuum (it gives the $W^3 B$ mass mixing) and so we know that, unlike
for the $T$ parameter, no global ({\em ergo} momentum-independent)
symmetry can make $S$ vanish. 

However, it is worth observing that $\Pi^\prime_{3B}$ transforms as a
triplet of $SU(2)_L$. $SU(2)_L$ is not a symmetry of the vacuum,
because it is broken the presence of the electroweak vev, $v$, which
is a doublet. This implies that (because for $SU(2)$ irreps, $2 \otimes 2 = 3
\oplus 1$) $S \propto v^2$. 

(Note that we are really making a spurionic argument, again.)

To recap, we have argued that $S \propto
v^2$. But $S$ is dimensionless, so $S \propto
v^2/f^2$, where $f$ is some dimensionful scale of the EWSB
dynamics. For technicolour, $f$ is the techni-analogue of $f_\pi$, and
$v \sim f$, so $S \sim 1$. But what if we could find a theory of
strong dynamics (so as to solve the hierarchy problem in the same way
as QCD) in which $v$ was somewhat smaller than $f$, either by accident
or design? This is where the composite Higgs enters the story.

To describe it in more detail, consider a different point of view. In
QCD and technicolour, the pattern of global symmetry breaking is
$SU(2)_L \times SU(2)_R \rightarrow SU(2)_V$ (times an overall $U(1)$
that we'll ignore for now). This is desirable because (i) we can embed
the SM gauge group in it and get the right pattern of breaking and
(ii) it naturally protects the value of the $T$ parameter. Now, we
have already seen that, at least locally, $SU(2) \simeq SO(3)$ and
$SU(2)\times SU(2) \simeq SO(4)$. This means that we can also write
the breaking as $SO(4) \rightarrow SO(3)$. The advantage of writing it
this way is that we can easily see how to change the symmetries,
while preserving the two desirable features just described: we can
have any $G \rightarrow H$ where $G$ contains $SO(4)$ and $H$ contains
$SO(3)$.

The first, obvious, extension is to consider $SO(5) \rightarrow SO(4)$
\cite{Agashe:2004rs} (though other examples have been considered
\cite{Gripaios:2009pe,Mrazek:2011iu}). The Lie algebra of $SO(n)$ is generated by antisymmetric,
imaginary matrices and has dimension $\frac{n(n-1)}{2}$. This means
that there are $10 - 6 = 4$ broken generators and so 4 Goldstone
bosons. Moreover, it is easy to show (exercise \ex) that those Goldstone
bosons transform as a $4$ of the unbroken $SO(4)$ subgroup. But
we have already seen that a $4$ of $SO(4)$ is a $(2,2)$ of $SU(2)\times SU(2)$,
which are exactly the charges of the SM Higgs doublet. To summarise:
a strongly-coupled model with $SO(5) \rightarrow SO(4)$ produces a set
of Goldstone bosons that have precisely the same charges as the SM
Higgs doublet, which of course is what we observe!

Our excitement is tempered somewhat by the realization that the Higgs
boson, with mass 125 GeV, looks nothing like a massless Goldstone
boson.
However, we know that the $SO(5)$ symmetry of our model cannot be
exact, because Nature manifestly does not exhibit it. For one thing,
the SM fermions cannot be arranged into degenerate multiplets of
the unbroken $SO(4)$. For another, we know that an $SU(2) \times
U(1)$ subgroup of $SO(5)$ is gauged, and this breaks $SO(5)$ by singling out
certain generators.

So we know the symmetry can only be at best approximate, in which case
the Higgs is at best an approximate, or {\em pseudo} Goldstone
boson. In particular, it will acquire a potential, and non-derivative
couplings to other particles, just like the SM Higgs. 

All of this can be computed explicitly in an EFT formalism for the
low-energy Goldstone bosons and gauge bosons (but just like in QCD, we
cannot compute the EFT parameters themselves from the underlying
strongly-coupled theory; we can only estimate their size using
na\"{\i}ve dimensional analysis). Such an EFT is
called a {\em non-linear sigma model}. Unfortunately, these are often
written down in a haphazard way in the literature, leading to results
that are not always correct. As an example, it is often claimed in
the literature that the contribution to the Higgs potential coming
from the gauge bosons in an
$SO(5) \rightarrow SO(4)$ model is given by
\begin{gather} \label{eq:inv1}
V(h) = A (3g^2 + g^{\prime 2}) \sin^2 \frac{h}{f},
\end{gather}
where $A$ is some constant and $f$ is the scale of $SO(5) \rightarrow
SO(4)$ breaking (it is the analogue of $f_\pi$ in QCD). In fact it is given by the more general
expression
\begin{gather}
V(h) = 2A (3g^2 \cos^4 \frac{h}{2f} +  g^{\prime 2} \sin^4 \frac{h}{2f} )
+ 2B  (3g^2 \sin^4 \frac{h}{2f} +  g^{\prime 2} \cos^4 \frac{h}{2f} ),
\end{gather}
which reduces (exercise \ex) to the claimed expression only when $A=B$.
In fact, there is a general result for $G \rightarrow H$, which is that one can write down
one invariant term in the lagrangian for each real or pseudo-real
irrep of $H$ that appears in decomposing the adjoint irrep of $G$, but
that one of these is an $h$-independent constant.
So (exercise \ex), for $SO(5) \rightarrow SO(4)$ the 10-d adjoint decomposes as $10
\rightarrow (3,1) + (1,3) + (2,2)$ and there are two invariant
lagrangian terms.\footnote{I should remark that if we instead have
  $SO(5) \rightarrow O(4)$ (which is desirable to protect the rate for
  $Z\rightarrow b\overline{b}$ decays \cite{Agashe:2006at}), then $(3,1) \oplus (1,3)$ is an irrep of
  $O(4)$ \cite{Gripaios:2014pqa}, and we get only the single invariant in (\ref{eq:inv1}).}
\footnote{I wish I had time to show you how to do all this properly, but I
don't. If you want to figure it out for yourself, a good place to start to learn how to write down the EFT
{\em comme il le faut} is \cite{Preskill:1990fr}. Next, make the gauge
coupling a spurion transforming as an adjoint of both $G$ and a copy
of the subgroup to be gauged (call it $K$). Finally, write $G\times K$
invariants built out of the gauge coupling spurion and the Goldstone bosons.}

The problem with a potential like (\ref{eq:inv1}) is that its minimum
is at $h=0$, meaning no EWSB.\footnote{This is true more generally,
  when we gauge a subgroup of the unbroken subgroup $H$.}
Salvation comes in the form of the couplings of the strong sector to the SM
fermions, which must also break $SO(5)$, and thus generate contributions to
$V(h)$.

These coupings must be present, because we know that the Higgs (which
is here part of the strongly coupled sector) couples to fermions (and
gives them mass after EWSB).
There are two ways in which we can imagine the couplings arising. The
first is much like the SM Yukawa couplings, in that the strong sector 
couples to fermion bi-linears. Schematically, 
\begin{gather} \label{old}
\mathcal{L} \supset \frac{q \mathcal{O}_h u^c
}{\Lambda^{d-1}} + \dots,
\end{gather}
 where $\mathcal{O}_h$ is some operator in the
strong sector of arbitrary dimension $d$ with the right quantum numbers to couple to SM
fermions. 

However, to this EFT lagrangian we should also add other operators that are compatible with the symmetries of the theory. Amongst these are
\begin{gather} \label{danger}
\mathcal{L} \supset \frac{q q q q}{\Lambda^{2}} + \Lambda^{4 - d'}\mathcal{O}_{h}^\dagger \mathcal{O}_{h}.
\end{gather}
The first of these is responsible for flavour changing neutral
currents; for these to be small enough, $\Lambda > 10^{3-5}$ TeV. But
then, in order to get a mass as large as that of the top from the
operator in (\ref{old}), we need to choose $d$ to be rather small: $d
\lesssim 1.2 - 1.3$ \cite{Luty:2004ye}. Next, we need to worry about
the second operator in (\ref{danger}). In order not to de-stabilize
the hierarchy, its dimension, $d'$, had better be greater than four,
rendering it irrelevant.\footnote{It is, perhaps, instructive to see
  how the hierarchy problem of the SM is cast in this language. There,
  $\mathcal{O}_{h}$ corresponds to the Higgs field $h$, with dimension
  close to unity, whilst $\mathcal{O}_{h}^\dagger \mathcal{O}_{h}$ is
  the Higgs mass operator, with dimension close to 2.} So what is the
problem? The limit in which $d \rightarrow 1$ corresponds to a free
theory (for which the operator $\mathcal{O}_h$ is just the Higgs field
$h$), and in that limit $d' \rightarrow 2d \rightarrow 2$. So in order to have an
acceptable theory, we need a theory containing a scalar operator
$\mathcal{O}_h$ (with the right charges) with a dimension that is
close to the free limit, but such that the theory is nevertheless
genuinely strongly-coupled, with the dimension of
$\mathcal{O}^\dagger_h \mathcal{O}_h$ greater than four. We have very
good evidence that such a theory cannot exist \cite{Rattazzi:2010yc}.

In the other approach, we imagine that the elementary fermions couple
linearly to fermionic operators of the strong sector \cite{Kaplan:1991dc}. Schematically, the lagrangian is
\begin{gather} \label{mix}
\mathcal{L} \sim   q \mathcal{O}_{q^c} +    u^c \mathcal{O}_{u} +   \mathcal{O}_{q^c} \mathcal{O}_{q} + \mathcal{O}_{u^c} \mathcal{O}_{u} +  \mathcal{O}_{q^c} \mathcal{O}_H \mathcal{O}_{u}
\end{gather}
(where I have left out the $\Lambda$s)
and the light fermion masses arise by mixing with heavy fermionic
resonances of the strong sector, which feel the electroweak symmetry
breaking. The beauty of this mechanism is that fermion masses can now
be generated by relevant operators ({\em cf.} the operator that
generates masses in (\ref{old}), which is at best marginal, since
$d>1$); this means that one can, in principle, send $\Lambda$ to
infinity and the problems with flavour physics can be
completely decoupled. There is even a further bonus, in that the light
fermions of the first and second generations, which are the ones that
flavour physics experiments have most stringently probed, are the ones
that are least mixed with the strong sector and the flavour-changing
physics that lies therein. In this model, the observed SM fermions are
mixtures of elementary and composite fermions, with the lightest
fermions being mostly elementary, and the top quark mostly
composite. The scenario therefore goes by the name of {\em partial
  compositeness}.

It turns out (see, {\em e.\ g.}, \cite{Contino:2010rs}) that the fermions can give negative corrections
to the mass-squared in the Higgs potential, and thus result in
EWSB. Since the top quark Yukawa is somewhat bigger than the gauge
couplings, this is (at least na\"{\i}vely) the most likely outcome.

We now have something approaching a realistic model of EWSB via strong
dynamics. Having built it up, we should now do our best to knock it
down. 

A first problem is that no one actually knows how to get a pattern of
$SO(5) \rightarrow SO(4)$ global symmetry breaking out of an explicit
strongly-coupled gauge theory coupled to fermions.\footnote{The
  breaking $SO(6) \rightarrow SO(5)$ \cite{Gripaios:2009pe} is
easier to achieve, since $SO(6) \simeq SU(4)$, and unitary
groups are easier to obtain.} 

A second problem is the $S$-parameter. We have argued that the
necessary suppression can be obtained if $v$ turns out to be somewhat
smaller than $f$, the scale of strong dynamics. Well, $v$ is obtained
by minimizing the Higgs potential $V(h)$, which contains contributions
of very roughly equal size, but opposite in sign, from the top quark
and gauge bosons. Thus it is possible to imagine that there is a
slight cancellation due to an accident of the particular strong
dynamics, such that the $v$ that emerges is small enough. A measure of
the required tuning is $\frac{v^2}{f^2}$, and the observed $S$-parameter requires
tuning at the level of ten {\em per cent} or so.

The third problem concerns flavour physics. To argue, as we have done above,
that the flavour problem can be decoupled, is not the same as arguing
that it {\em is} solved. To do that, one
needs to find an explicit model which possesses all the required
operators, with the right dimensions. Needless to say, our ignorance
of strongly-coupled dynamics means we have no idea whether such a
model exists. Certainly, in all cases that have been studied (either
models with large rank of the gauge group, or lattice studies), there
{\em is} a problem with flavour constraints. 

Note that in both cases, we have an `advantage' with respect to
technicolour models, in that we can always suppose that the peculiar
dynamics of the model makes $v$ somewhat smaller than $f$, whereas in
technicolour we are stuck with $v \sim f$. Still we should remember
that it is precisely to avoid having to do this kind of tuning that we
built such models in the first place!

Despite these problems, composite Higgs models seem just as good (or
just as bad) as solutions to the hierarchy problem as supersymmetric
models, and so they deserve thorough investigation at the LHC. This
itself is not so easy to do. Na\"{\i}vely, the obvious place to look
for deviations is in the Higgs sector itself, for example in the
couplings of the Higgs boson to other particles. However, we know that
(since such models reproduce the SM in the limit $v^2/f^2 \rightarrow
0$) the deviations must be proportional to $v^2/f^2$ and hence at
most 10 \% or so. Such deviations are hard to see at the LHC,
and even at a future $e^+ e^-$ collider. Perhaps a better way is to
look for the composite partners of the top quark, which must be not
too heavy in order to reproduce the observed Higgs mass. Many
suggestions for how to do so have been put forward and the experiments
are beginning to implement them. See, {\em e.g.} \cite{DeSimone:2012fs,Gripaios:2014pqa} and
refs. therein for more details.

\section{The axion}
The axion offers an elegant solution to a mysterious problem of the
SM called the `strong $CP$ problem'. It is, moreover, an excellent
candidate for the dark matter that makes up 20 \% of the energy
density of the Universe today. It behoves us to discuss it. 

We begin by rectifying some earlier sleight of hand. In introducing
the composite Higgs, we claimed that strong-coupling in QCD breaks the
$SU(2)_L \times SU(2)_R$ (approximate) chiral symmetries to the
vectorial $SU(2)_V$, resulting in 3 (nearly) massless pions. But the
chiral symmetry is really $U(2)_L \times U(2)_R$ which gets broken to
$SU(2)_V \times U(1)_B$, so why isn't there a fourth pion,
corresponding to the axial $U(1)$? This old mystery
(which went by the name of the $U(1)_A$ problem) is explained by
the fact that the $U(1)_A$ is not a symmetry, because of a quantum
{\em anomaly}. (If you don't know about these already, you'd better
look in a QFT textbook now.) For our purposes, it is enough to note that
rotating
$(q,q^c) \rightarrow e^{i\theta} (q,q^c)$, does not send the
QCD lagrangian to itself, but rather generates the term
\begin{gather} \label{eq:thetaterm}
\frac{\theta g_s^2}{16 \pi^2} F^a_{\mu \nu} \tilde{F}_a^{\mu \nu},
\end{gather}
where $\tilde{F}_a^{\mu \nu} \equiv \frac{1}{2} \epsilon^{\mu \nu \sigma \rho}
F_{a \sigma \rho}$. Now,  we omitted this term from our
original gauge lagrangian (\ref{eq:lgauge}) because it may be written
as a total derivative and therefore does not contribute at any order
in perturbation theory. But it does not vanish non-perturbatively (in
particular, it gets contributions from instanton
configurations). Thus, there is no symmetry to break, and no fourth
Goldstone boson. The $U(1)_A$ problem is solved. 

In its place, we find a different problem. The term
(\ref{eq:thetaterm}) violates $CP$. If all quarks are massive, then we
cannot completely rotate it away using the above chiral rotation, and
indeed we find that the combination $\overline{\theta} = \theta +
\mathrm{arg\; det} M$ is physical, where $M$ is the quark mass
matrix. Physical quantities like the vacuum energy and the electric
dipole moment (EDM) of the neutron will depend on $\overline{\theta}$. These can be
computed using the EFT for the pions (called {\em chiral perturbation
  theory}; for a review, see \cite{Colangelo:2000zw}); the vacuum energy density, for example, is given by
\begin{gather} \label{eq:thetaterm}
E(\theta) = -m_\pi^2 f_\pi^2 \frac{m_u m_d}{(m_u + m_d)^2} \cos^2 \overline{\theta}.
\end{gather}

Measurement (or rather lack of) a neutron EDM leads to a bound of
$\overline{\theta} \lesssim 10^{-9}$. Since $\overline{\theta}$ is just an angular
parameter (and a combination of strong and weak sector physics at that,
with the latter already known to contain a distinct, $O(1)$,
$CP$-violating phase), a na\"{\i}ve guess for its size would be $O(1)$. The strong
$CP$ problem is to understand why it is, in fact, so small. 

The axion provides a compelling solution. To introduce it, note that
if we could somehow turn $\overline{\theta}$ into a field, then the
vacuum energy (\ref{eq:thetaterm}) would act as a potential for the
field, which we should minimize to obtain $\overline{\theta} = 0$. We
can achieve this miracle by making the chiral rotation of quarks a symmetry of the
lagrangian, in the absence of the anomaly \cite{Peccei:1977hh}. This requires extra
fields. The original way of doing it (due, independently, to Weinberg
\cite{Weinberg:1977ma} and Wilczek \cite{Wilczek:1977pj}),
for example, was
to imagine that there are 2 Higgs doublets, one of which is
responsible for up quark masses and the other for down quark masses
(like in SUSY). The corresponding Yukawa terms are
\begin{gather} \label{eq:2hdm}
\lambda^u q H_u u^c + \lambda^d q H_d d^c + \mathrm{h.\ c.}
\end{gather}
and it is clear that the chiral rotations of the quarks can now be
undone (modulo the anomaly) by equal and opposite transformations of
$H_u$ and $H_d$. At this point, we might be tempted to argue that, just as if we
had a massless quark, we could remove $\overline{\theta}$ by an
anomalous symmetry transformation, making it unphysical. 

But we cannot do so, because the extra symmetry must be spontaneously
broken in the vacuum. Indeed, the symmetry acts chirally on quarks,
and we know that this must be broken in their vacuum by their
masses. So (in the absence of the anomaly) there is a Goldstone
boson $a(x)$, called the axion.
The axion is a spacetime dependent field, and we cannot remove it
by a constant rephasing.
The net effect is that
$\overline{\theta}$ is replaced everywhere by the dimensionless combination
$a(x)/f_a$ (up to a constant, which we {\em can} re-phase away), where $f_a$ is the symmetry breaking order parameter (just
like $f_\pi$ for pions or $f$ for the composite Higgs). Note that,
because of the anomaly, $a(x)$ is not really a Goldstone boson. It
gets a potential $V(a) = -m_\pi^2 f_\pi^2 \frac{m_u m_d}{m_u + m_d}
\cos^2 \frac{a}{f_a}$ from (\ref{eq:thetaterm}), which is minimised at
$a=0$, solving the strong CP problem. It also acquires couplings to
other matter, which are proportional to $1/f_a$. 

So all the phenomenology is fixed by $f_a$ and we can translate
experimental data into bounds on it. In the original Weinberg-Wilczek
model, the symmetry is broken by the Higgs vevs, which are also
responsible for EWSB, and so
$f_a$ is of order $v$. This model was very
quickly ruled out, but it was soon realised that $f_a$ can be given
whatever value one likes, by making the axion symmetry breaking scale
independent of the weak scale. This is called the `invisible axion' scenario.

What then are the observational constraints on $f_a$ (see, {\em e.\ g.},
\cite{Turner:1989vc,Raffelt:1990yz} for more details)? If $f_a$ is
below about $10^9$ GeV, the axion is sufficiently strongly coupled to
other matter (whilst remaining sufficiently light) to be able to
compete with neutrini in transporting energy out of cooling stars. In
particular, there would have been observational consequences in {\em
  e.\ g.}, the 1987 supernova and red giants.

We can get an upper bound as follows. On cosmological scales, the
axion field is Hubble damped until $t^{-1} \sim m_a$. There is no
reason to suppose that the axion sits at its minimum during this
period.
Subsequently, for $t^{-1} \lesssim m_a$, the axion undergoes damped
oscillations about its minimum with the amplitude being diluted by the expansion.
It thus behaves like cold dark matter.\footnote{If the symmetry
breaking that yields the axion occurs after inflation, then there
should also be defects in the axion field in the cosmos that
contribute to dark matter, making the bounds more model dependent.} 
As $f_a$ is increased, the oscillations start later and are diluted less,
and so we end up with too much dark matter today. 
If $f_a \gtrsim 10^{12}$ GeV, the Universe is overclosed, but if
not, then the axion is an excellent dark matter
candidate. Unfortunately, the axion is so weakly coupled to matter in
the allowed window of $f_a$ that it is very difficult to detect, but
several ingenious experiments have either been proposed or are already
under way, {\em e.\ g.} \cite{Asztalos:2009yp}. I encourage you to read more about them. 
\section{Afterword}
I have only had time to discuss one or two examples, in the briefest
terms, but I hope it is clear to you that there is plenty of evidence
for physics BSM, and plenty of ideas for what constitutes it. This is
good news. The bad news is that a lot of that evidence (the
cosmological constant, neutrino masses, GUTs, inflation, the axion,
quantum gravity) hints at
scales that are a long, long way away from our current high-energy
frontier ({\em c.} TeV). Of course, the electroweak hierarchy problem
is still a very strong motivation for probing the TeV scale, but if
nothing turns up in the next LHC run, then you (and your experimental
colleagues) are going to need to get a lot more creative in the
future. I wish you the best of luck!
\providecommand{\href}[2]{#2}\begingroup\raggedright\endgroup

\end{document}